\documentclass[12pt]{iopart}
\input{iopams.sty}
\usepackage{graphicx}
\usepackage[dvips]{color}

\newcommand{\Slash}[1]{\ooalign{\hfil/\hfil\crcr$#1$}}
\begin{document}

\title[]{Density fluctuations and chiral phase transition}

\author{K Redlich$^{1,2}$, B Friman$^2$,  C Sasaki$^3$}

\address{$^1$ Institute of Physics
Theoretical Physics, University of Wroc\l aw, 50204 Wroc\l aw, PL}
\ead{redlich@ift.uni.wroc.pl}
\address{$^2$ Gesellschaft f\"ur Schwerionenforschung, GSI,  D-64291 Darmstadt,
D} \ead{b.friman@gsi.de}
\address{$^3$ Technische Universit\"at M\"unchen,  D-85748 Garching, D }
 \ead{csasaki@ph.tum.de}
\begin{abstract}
Based on an effective QCD Lagrangian we discuss the properties of charge density
fluctuations in the vicinity of chiral phase transition. We explore thermodynamics in the
presence of spinodal phase separation.  We show that appearance of   spinodal
decomposition in a non-equilibrium first order phase transition results in  divergence of
the charge density fluctuations related with the electric charge and baryon number
conservation. Consequently, divergent fluctuations at the chiral phase transition  are
not only attributed to the critical  end point but are also there along the first
order phase transition  if the spinodal phase separation take place. Based on the mean
field dynamics, the critical exponents for these singular behavior of charge
susceptibilities are also discussed.
\end{abstract}

\maketitle

\section{Introduction}

One of the objectives of  experiments with ultra-relativistic heavy ion collisions is to
map the QCD phase diagram and study the properties of high density strongly interacting
medium. The recent results obtained within Lattice Gauge Theory (LGT) description of  QCD
thermodynamics  at vanishing baryon density show that the phase transition from hadronic
phase to quark-gluon plasma is a  cross-over  \cite{lgt}. From effective chiral models
calculations \cite{kunihiro,ef5,ising,sexy,pd,fujii,stephanov,hatta1,SFR:NJL,hatta,bj} as
well as from the first LGT studies at finite baryon density \cite{dlgt} one concludes
that at finite chemical potential the cross-over transition is most likely to be
converted to the first order. Consequently, there might exist  a critical end point
(CEP) on the QCD phase diagram as a matching point of first order and cross-over
transition. Based on the universal properties of the QCD chiral phase transition, one
also expects, that the CEP belongs to the 3-dimensional Ising model universality class
\cite{hatta,ejiri}. This implies, that the CEP is a particular point on the QCD phase
diagram where the fluctuations of the net quark and electric charge densities are
diverging \cite{stephanov,hatta,bj}.

In heavy ion phenomenology the fluctuations of conserved charges are directly accessible
experimentally. Thus, a non-monotonic behavior of such fluctuations in the c.m.s
collision energy in heavy ion experiments would be an ideal and transparent signal of the
existence of the  CEP \cite{SFR:NJL}. However, such conclusion is based on the assumption
that the first order phase transition appears in equilibrium. In heavy ion collisions, we
are dealing with quickly expanding dynamical system, such that, any deviation from
equilibrium are not excluded.

A first-order phase transition is intimately linked with the existence of a convex anomaly in
the thermodynamic pressure \cite{ran,gavin,rans,ranshi,SFR:spinodal}. There is an
interval of energy density or baryon number density where the derivative of the pressure
is positive. This anomalous behavior characterizes a region of instability in the
temperature, T and baryon density, $n_q$ plane. This region is bounded by the spinodal
lines, where the pressure derivative with respect to volume vanishes. The above anomalous
properties of the first order transition can be uncovered in non-equilibrium system.

In this work we discuss a possible  influence  of the spinodal phase separation on the
properties of baryon number density fluctuations. We show that spinodal instabilities
result in divergence of electric charge $\chi_Q$ and baryon densities $\chi_{\mu\mu}$
fluctuations. Consequently, a critical behavior of charge fluctuations  is not only
attributed to the CEP but is also there along the first order transition if spinodal phase
separation appears in a medium.

Our discussion will be based on the effective chiral model calculations, however our main
conclusion on the properties of charge fluctuations in the presence of spinodal
instability is quite general and is independent on the particular choice of the effective
chiral Lagrangian.

 \setcounter{equation}{0}
\section{Effective Chiral  Model and Spinodal instabilities}
\label{sec:NJL}

To study the properties of  the chiral phase transition in the presence of spinodal
instabilities we use as an effective chiral Lagrangian  the Nambu--Jona-Lasinio (NJL)
model~\cite{nambu}. This  model describes the interactions of quarks preserving the
chiral symmetry of the massless QCD Lagrangian.  To consider thermodynamics we constrain
the quark interactions only to the scalar and isoscalar sectors that are controlled by the
phenomenological coupling $G_S$. For two quark flavors, three colors and for finite quark
chemical potentials the Lagrangian has a well known structure
~\cite{kunihiro,sexy,review}:
\begin{equation}\label{eq1}
{\mathcal L} = \bar{\psi}( i \Slash{\partial} -m )\psi {}+ \bar{\psi}\mu_q\gamma_0\psi {}+ G_S
\Bigl[ \bigl( \bar{\psi}\psi \bigl)^2 + \bigl( \bar{\psi}i\vec{\tau}\gamma_5\psi \bigl)^2
\Bigr]\,,
\end{equation}
where $m = \mbox{diag}(m_u, m_d)$ is  the current quark mass, $\mu_q = \mbox{diag}
(\mu_u, \mu_d)$ is the  quark chemical potential  and $\vec{\tau}$ are Pauli matrices.
The momentum cut-off and the strength of the quark interaction are fixed such that
to reproduce the vacuum pion parameters.

\begin{figure*}
\begin{center}
\includegraphics[width=7.7cm]{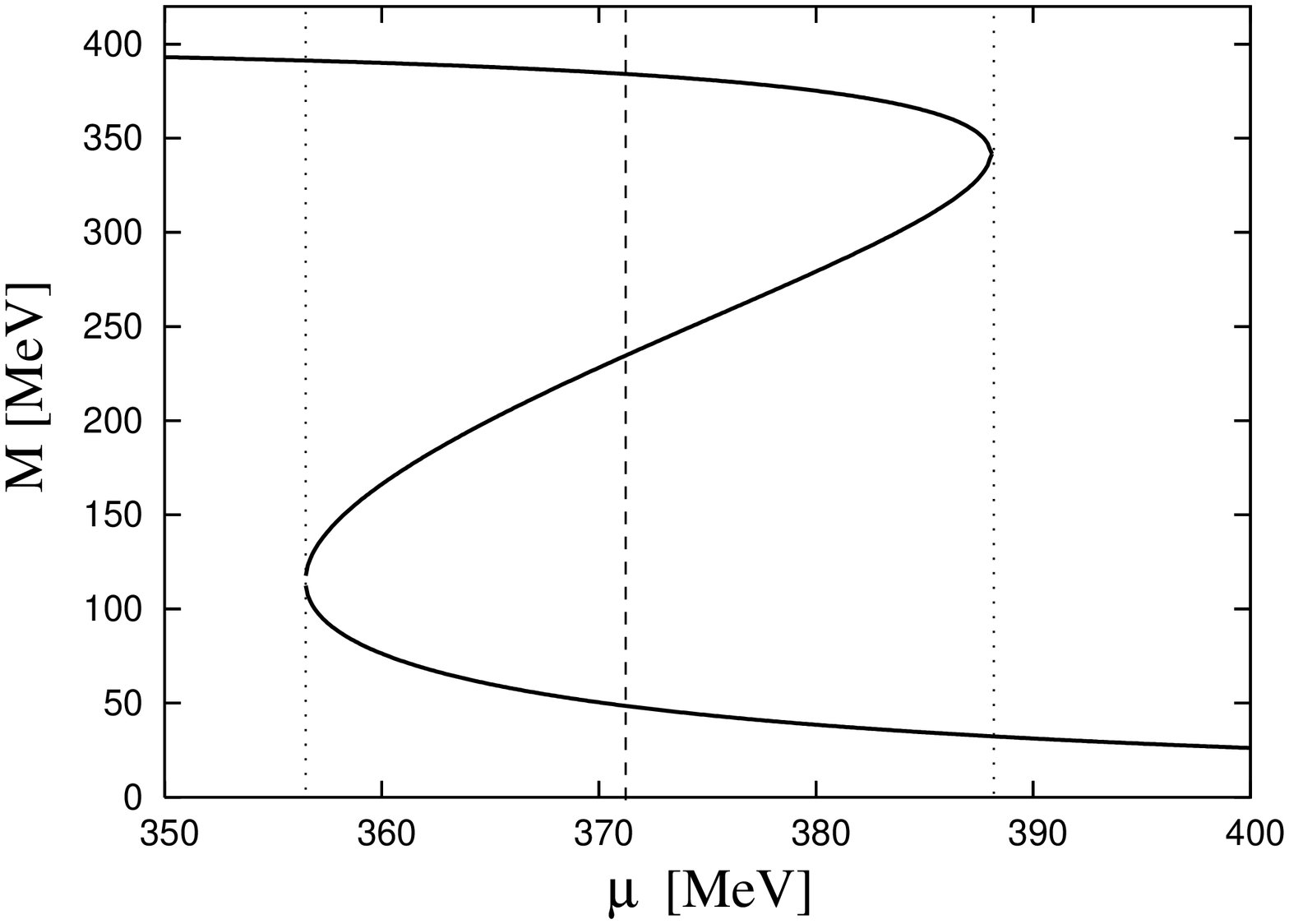}
{\includegraphics[width=7.8cm]{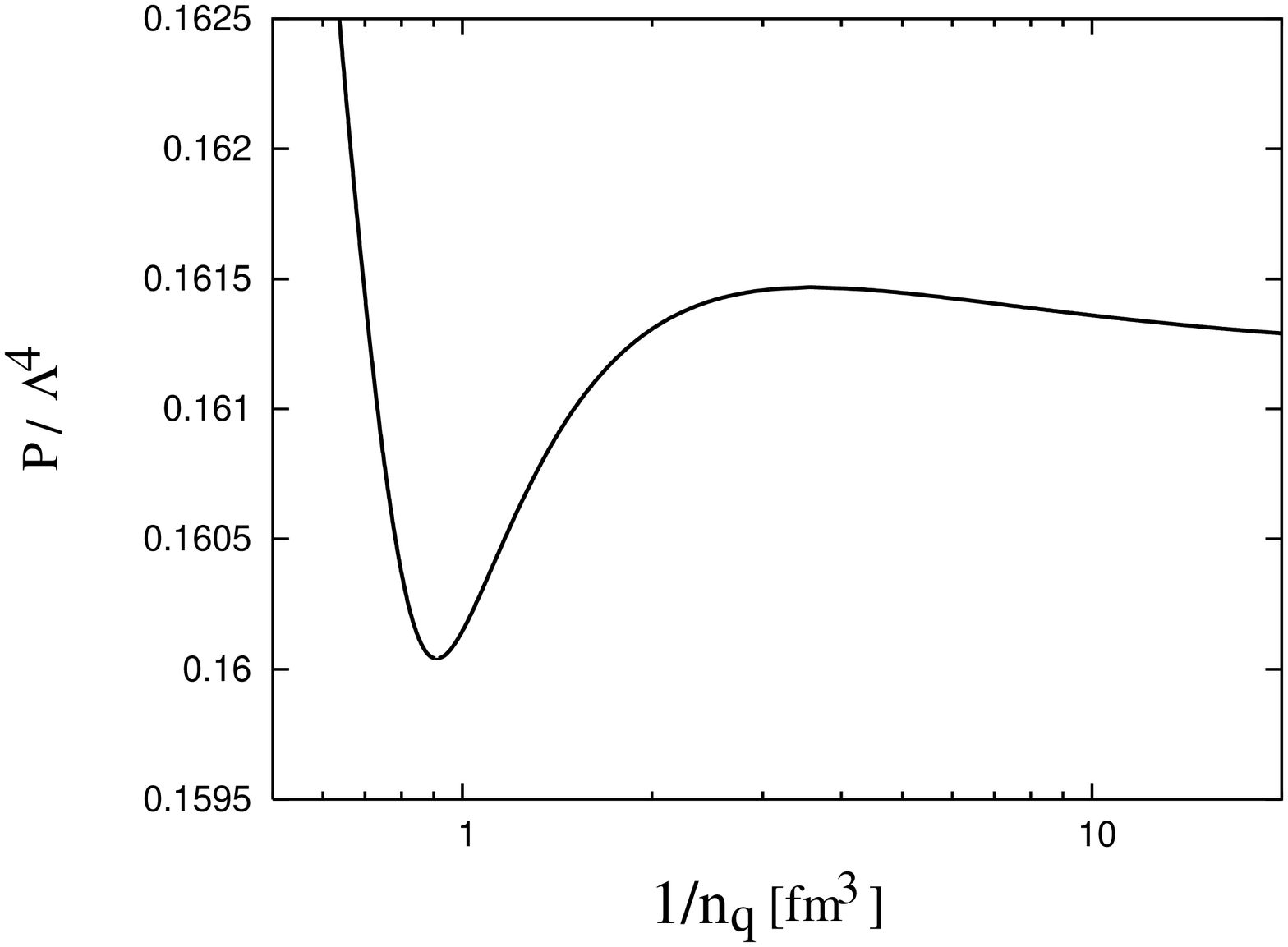}} \caption{ The left hand figure: the dynamical
quark mass at fixed temperature as a function of quark chemical potential. The broken
line indicates an equilibrium first order phase transition. The dotted lines constrain
the isothermal spinodal points. The right hand figure: the pressure as a function of
inverse quark number density for fixed temperature $T=30$ MeV.\protect\cite{ournew}
 }
\label{fig1}
\end{center}
\end{figure*}

%

To study the thermodynamics of the NJL model we use the mean field approximation. For an
isospin symmetric system the thermodynamic potential  is obtained as~\cite{review}:
\begin{eqnarray}\label{eq2}
& \Omega (T,\mu;M)/V = \frac{(M-m)^2}{4G_S} {}- 12 \int\frac{d^3p}{(2\pi)^3}
\Bigl[E(\vec{p}\,)
\nonumber\\
& {}- T\ln ( 1-n^{(+)}(\vec{p},T,\mu) )\Bigr. {}-\Bigl.  T\ln (1-n^{(-)}(\vec{p},T,\mu)
\Bigr]\,, \end{eqnarray} with  $M = m- 2G_S\langle \bar{\psi}\psi \rangle$ being a
dynamical quark mass, $E(\vec{p}\,) = \sqrt{\vec{p}^{\,2} + M^2}$  its energy and
 $n^{(\pm)}(\vec{p},T,\mu) = \Bigl( 1
+ \exp\bigl[ (E(\vec{p}\,) \mp \mu)/T \bigr] \Bigr)^{-1}$ are the particle/antiparticle
distribution functions. The quark chemical potential $\mu$ is expressed as an average
$\mu=(\mu_u +\mu_d)/2$.

The dynamical quark mass $M$ in  Eq. (\ref{eq2}), that places a role of the order
parameter of chiral phase transition, is obtained self-consistently from the stationarity
condition ${\partial\Omega}/{\partial M} = 0$ that leads to:
\begin{equation}\label{eq3}
M = m+24 G_S \int\frac{d^3 p}{(2\pi)^3} \frac{M}{E} \Bigl[ 1 - n^{(+)} - n^{(-)}
\Bigr]\,.
\end{equation}
  In Fig.~\ref{fig1}-left the constituent quark mass is shown  at fixed temperature
$T=30$ MeV  for different values  of quark chemical potentials. The behavior of $M$, seen
in Fig.~\ref{fig1}, is typical for systems that exhibit a first order phase transition:
There is no unique solution of the gap equation. Instead, there are metastable solutions
that correspond to the local minima of thermodynamic potential. For  finite current quark
masses in the NJL Lagrangian the chiral symmetry is explicitly broken. Consequently, $M$
is not any more an order parameter and is never zero as seen in Fig.~\ref{fig1}.

To identify the equilibrium transition from chirally broken to approximately symmetric
phase one usually  performs the Maxwell construction. In this case the chiral phase
transition parameters are fixed such that the three extrema of the thermodynamic
potential are degenerate.    The location of an equilibrium transition  from massive
quasiparticles to almost massless quarks, calculate at fixed $T=30$ MeV, is shown as
dashed-line in Fig.~\ref{fig1}.

\begin{figure}
\begin{center}
\includegraphics[width=8.7cm]{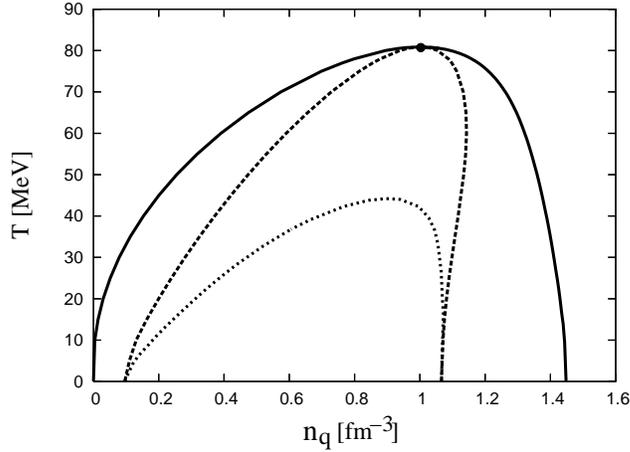}
\caption{  The phase diagram in the temperature $T$ and quark number  density $n_q$ plane
 in the NJL model. The filled point indicates the CEP.
The full lines starting at the  CEP represent  boundary of the mixed phase in
equilibrium. The broken-curves are the isothermal whereas the dotted ones are the
isentropic spinodal lines.
 }
 \label{fig2}
\end{center}
\end{figure}

  The non-monotonic behavior of the dynamical quark mass $M$ seen in Fig. \ref{fig1}
affects any thermodynamic observable since $M(T,\mu_q)$
 determines the properties of  medium constituents.
In particular, the thermodynamic pressure is directly obtained from the potential,
 $P=-\Omega$, whereas  the net quark number density
\begin{eqnarray}\label{eq4}
n_q &= 12 \int\frac{d^3 p}{(2\pi)^3} \Bigl[ n^{(+)} - n^{(-)} \Bigr]\,.
\end{eqnarray}
Thus, in the mean field approximation the   density $n_q$ has the same  structure as in a
non-interacting gas of massive fermions, however with the $(T,\mu)$-dependent effective
quark mass $M$.

Fig.~\ref{fig1}-right  shows the inverse density dependence of thermodynamic pressure at fixed
temperature. The pressure exhibits a non-monotonic structure as a consequence of the
properties of the dynamical quark mass seen in Fig.~\ref{fig1}-left. The unstable solution
of the gap equation leads to  mechanical instabilities in the thermodynamic pressure
where its volume derivative is positive. This region appears between spinodal points
characterized by the minimum and the maximum of the pressure. Outside of this region the
system is mechanically stable. The volume dependence of the pressure can be studied at
fixed temperature or at  fixed entropy. In the first case the spinodal points are
isothermal whereas in the second they are isentropic. Changing the temperature or entropy
results as the isothermal or isentropic spinodal lines in ($T,n_q)$-plane. If the volume
derivative of $P$ exists then the spinodal lines can be defined through the following
conditions:
\begin{equation}\label{eq5}
\left( \frac{\partial P}{\partial V}\right)_T=0 \qquad {\rm or} \qquad \left(
\frac{\partial P}{\partial V}\right)_S=0\,.
\end{equation}

\begin{figure}
\begin{center}
\includegraphics[width=7.2cm]{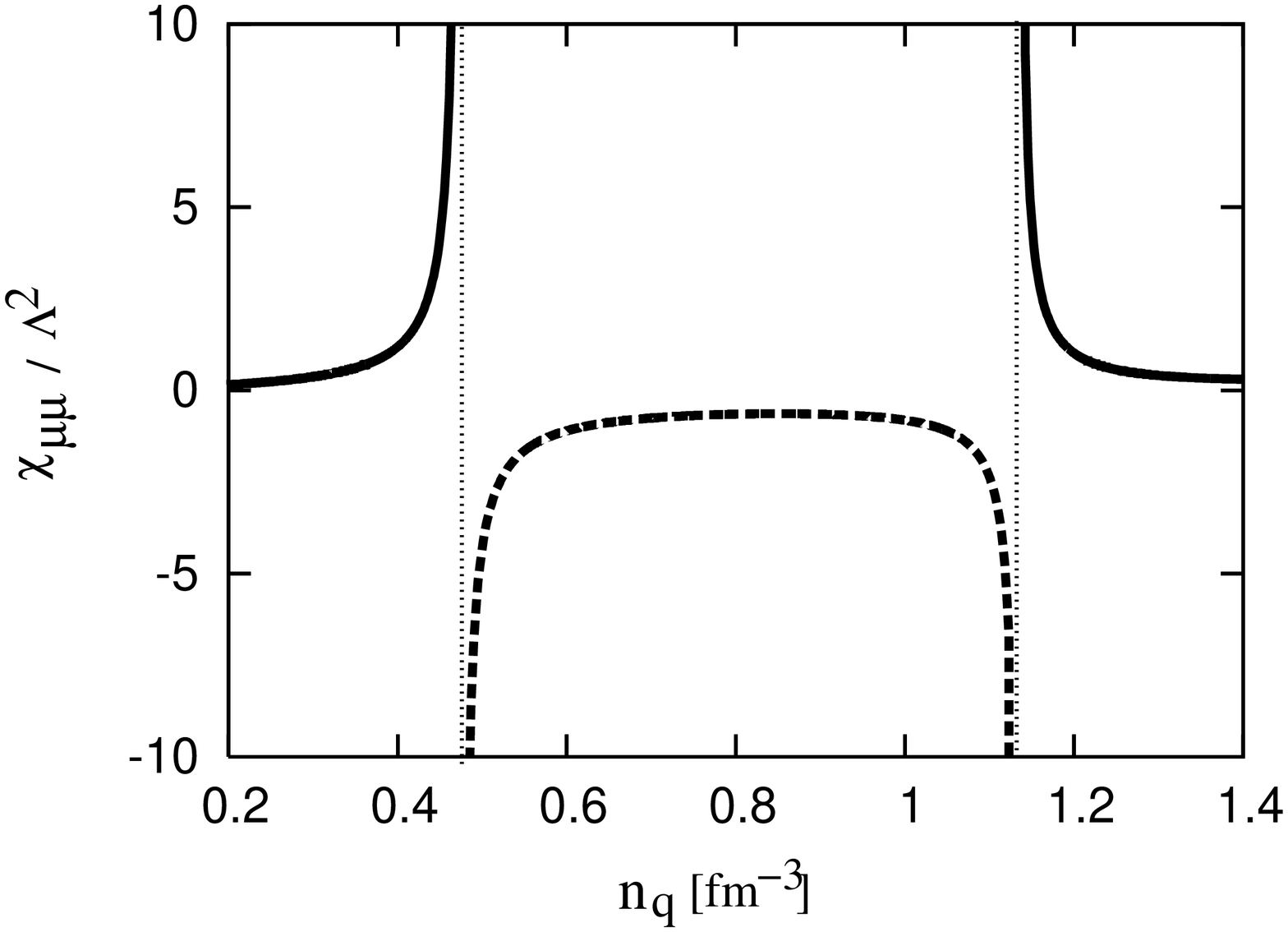}
\includegraphics[width=8.3cm]{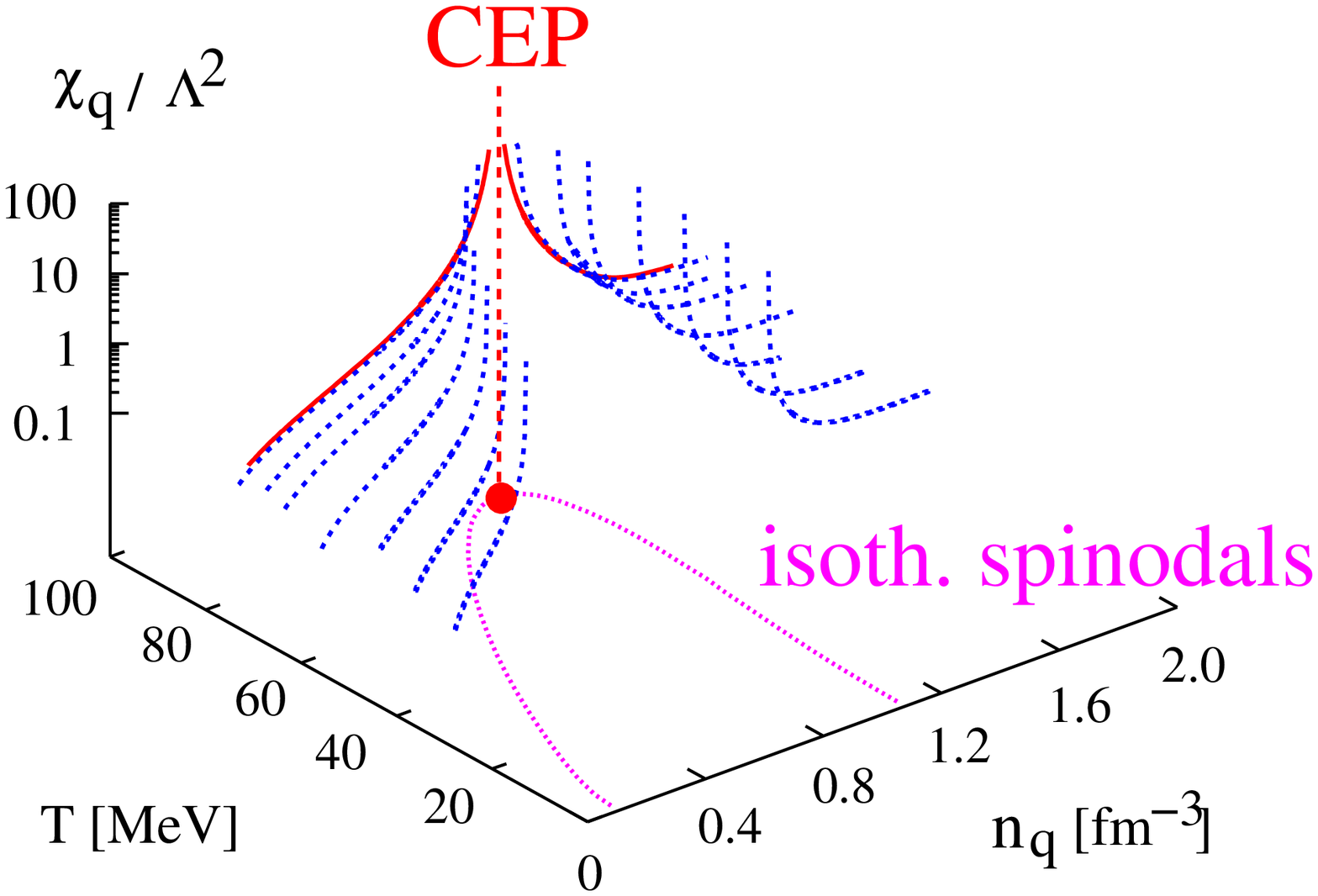}
\caption{ (Left) The net quark number susceptibility at $T=50$ MeV as a function of the
quark number density across the first order phase transition. (Right) The net quark
number susceptibility in the stable and meta-stable regions~\cite{SFR:spinodal}. }
\label{fig3}
\end{center}
\end{figure}

Considering  behavior of the dynamical quark mass and thermodynamic pressure  one can
 find the phase
diagram related with the chiral phase transition. Fig.~\ref{fig2} shows the resulting
diagram in the NJL model in the $(T,n_q)$-plane that accounts for spinodal instabilities.
The NJL model yields a generic, QCD like, phase diagram. It exhibits a CEP
that separates cross over from the first order chiral phase transition.

Assuming equilibrium transition there is a coexistence phase that ends at the CEP.
However, accounting  for expected instabilities due to a convex anomaly one
can distinguish the metastable  from mechanically unstable regions that are separated by
the spinodal lines. From the thermodynamic relation
\begin{eqnarray}\label{eq6}
& & \left( \frac{\partial P}{\partial V} \right)_T = \left( \frac{\partial P}{\partial V}
\right)_S {}+ \frac{T}{C_V}\left[ \left( \frac{\partial P}{\partial T} \right)_V
\right]^2\,,
\end{eqnarray}
it is clear that the isentropic spinodal lines are  lying inside the isothermal spinodals
and  that both lines coincide at $T=0$. In the mean field approximation the isothermal
spinodals  are matched with the cross over transition at the CEP whereas isentropic
spinodals  appear below CEP. The last property is modified when including quantum
fluctuations due to  changes in critical exponents of the specific heat with
constant volume $C_V$~\cite{SFR:spinodal}.
In the mean field approximation the $C_V$ is finite, whereas it
diverges in the quantum system. Consequently, from Eq. (\ref{eq6}) the isothermal and
isentropic conditions are equivalent at the CEP.

\begin{figure*}
\begin{center}
\includegraphics[width=8.7cm]{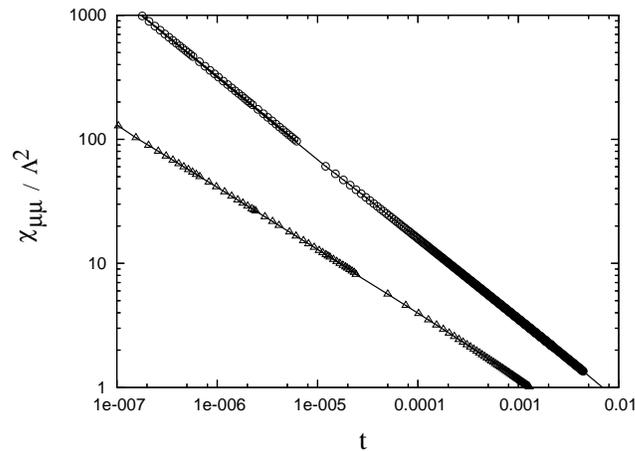}
\caption{The net quark number susceptibility in the vicinity of the CEP  as a function of
the reduced quark chemical  potential $t=(\mu-\mu_c)/\mu_c$ at fixed $T$. The  circle
denotes the results calculated at $T=T_{\rm CEP}$ and the  triangle  at $T=30$ MeV $<
T_{\rm CEP}$ corresponding to the first order transition.\protect\cite{ournew} }
\label{fig4}
\end{center}
\end{figure*}

\section{Charge fluctuations in the presence of spinodal phase separation}
We have already mentioned in the introduction that fluctuations of conserved charges are
excellent observables to study the critical properties in QCD medium related with chiral
phase transition. In statistical physics fluctuations of conserved charges are quantified
by corresponding susceptibilities,  $\chi_{\mu\mu}={\partial n_q}/{\partial\mu}$. Thus,
fluctuations of net quark number density $\chi_{\mu\mu}$ measures response of the density
to the change in quark chemical potential.

 Fig. ~\ref{fig3}-left   shows the evolution of the net
quark number fluctuations along a path of fixed $T=50$ MeV  in the $(T,n_q)$--plane. When
entering the coexistence region, there is a singularity in $\chi_{\mu\mu}$ that appears
when crossing the isothermal spinodal lines, where the fluctuations diverge and the
susceptibility changes sign. Between the spinodal lines, the susceptibility is negative.
This implies an instability of the baryon number fluctuations when crossing the
transition between the chirally symmetric and broken phases.

The behavior of $\chi_{\mu\mu}$ seen in Fig.~\ref{fig3}  is a direct consequence of the
thermodynamic relation
\begin{eqnarray}\label{eq7}
& \left( \frac{\partial P}{\partial V} \right)_T = -
\frac{n_q^2}{V}\frac{1}{\chi_{\mu\mu}}\,,\label{eq8}
\end{eqnarray}
which connects the pressure derivative with the susceptibilities. Along the isothermal
spinodal lines the pressure derivative in (\ref{eq7}) vanishes. Thus, for non-vanishing
density $n_q$, $\chi_{\mu\mu}$ must diverge to satisfy (\ref{eq7}). Furthermore, since
the pressure derivative ${\partial P}/{\partial V}|_T$ changes sign when crossing the
spinodal line, there must be a corresponding sign change in $\chi_{\mu\mu}$, as seen in
Fig.~\ref{fig3}-left.
Indeed, the negative specific heat in low energy nuclear collisions has been reported
as the first experimental evidence for the liquid-gas phase transition~\cite{experiment}.
Due to the linear relation between $\chi_{\mu\mu}$, the isovector
susceptibility $\chi_I$ and the charge susceptibility $\chi_Q$~\cite{stephanov,SFR:NJL},
the charge fluctuations are also divergent at the isothermal spinodal line. Thus, in
heavy-ion collisions, fluctuations of the baryon number and electric charge could show
enhanced fluctuations, as a signal of the spinodal decomposition
\cite{SFR:spinodal,ournew}. The spinodal phase separation can also lead to fluctuations
in strangeness \cite{rans} and isospin densities.

In the case of an equilibrium first order phase transition, the density fluctuations do
not diverge. The fluctuations
increase as one approaches the CEP along the first order transition and decrease again in
the cross over region. This led to the prediction of a non-monotonous behavior of the
fluctuations with increasing beam energy as a signal for the existence of a
CEP~\cite{stephanov,SFR:NJL}. We stress that strictly speaking this is relevant only for
the idealized situation where the first order phase transition takes place in
equilibrium. In the more realistic non-equilibrium system one expects fluctuations in a
larger region of the phase diagram, i.e., over a broader range of beam energies, due to
the spinodal instabilities \cite{SFR:spinodal}.

The singular behavior of the net quark number susceptibility at the CEP and isothermal
spinodals can be quantified by the corresponding critical exponents. Fig.~\ref{fig4}
shows the $\chi_{\mu\mu}$ calculated in the NJL model in the vicinity of the CEP and
spinodals as a function of the reduced quark chemical potential $t=(\mu-\mu_c)/\mu_c$ at
fixed temperature $T=30$MeV. There is a clear scaling of $\chi_{\mu\mu}\sim t^{-\gamma}$,
with the universal critical exponents $\gamma$. At the isothermal spinodal line this
exponent is found to be $\gamma=1/2$, while $\gamma=2/3$ at the CEP in agreement with the
mean-field results~\cite{hatta,bj}. Thus, the singularities at  spinodals yield a
somewhat stronger divergence as they join at the CEP \cite{ournew}.

 The mean field values of these
critical exponents extracted from the numerical studies shown in Fig.~\ref{fig4} can be
also derived analytically from the Ginzburg-Landau potential \cite{ournew} that is a good
approximation of any   effective chiral models   in a vicinity of chiral phase transition.

The critical exponents shown in Fig. \ref{fig4} are renormalized when including  quantum
and thermal fluctuations, but the smooth evolution of the singularity from the spinodal
lines to the CEP, illustrated in Fig.~\ref{fig3}-right, is expected to be generic being model
independent.

\section{Conclusions}

We have shown that the net quark number fluctuations  diverge at the isothermal spinodal
lines of the first order chiral phase transition. As the system crosses this line, it
becomes unstable with respect to spinodal decomposition. The unstable region is in
principle reachable in non-equilibrium systems, created e.g. in heavy ion collisions.
This means that large fluctuations of the baryon or electric charge density are expected
not only at the second order CEP but also at a non-equilibrium first order
transition.
\section*{ Acknowledgments}
 K.R. acknowledges partial support of
the Gesellschaft f\"ur Schwerionenforschung (GSI)   and the Polish Ministry of National
Education (MEN).

\end{document}